\documentclass[aps,twocolumn,floatfix]{revtex4} % showpacs

\usepackage{amsmath,amsfonts,amssymb,graphicx,times,color,bbm}

\newcommand{\me}{\mathrm{e}}
\newcommand{\mi}{\mathrm{i}}

\definecolor{green}{rgb}{0.2,.7,0.4}

\begin{document}

\title{Exploring local quantum many-body relaxation 
by atoms in optical superlattices}

\author{M.\ Cramer$^{1}$, A.\ Flesch$^2$, I.\ P.\ McCulloch$^3$,  
U.\ Schollw{\"o}ck$^2$, and J.\ Eisert$^{1,4}$} 

\affiliation{
1 Blackett Laboratory, 
Imperial College London,
London SW7 2BW, UK\\
2 Institut f{\"u}r Theoretische Physik C, 
RWTH Aachen University, 52056 Aachen, Germany\\
3 School of Physical Sciences, The University of 
Queensland, Brisbane, QLD 4072, Australia\\
4 Physics Department, University of Potsdam, 14469 Potsdam,
Germany}
   
\begin{abstract}
We establish a setting---atoms in
optical superlattices with period $2$---in which one can experimentally
probe signatures of the process of local relaxation and apparent thermalization in non-equilibrium dynamics without the need of addressing single sites. This opens up a way to 
explore the convergence of subsystems to maximum entropy 
states in quenched quantum many-body systems with
present technology. Remarkably, the emergence of thermal states does not
follow from a coupling to an environment, but is a 
result of the complex non-equilibrium dynamics in closed systems. 
We explore ways of measuring the relevant
signatures of thermalization in this analogue
quantum simulation of a relaxation process, 
exploiting the possibilities offered by optical superlattices.
\end{abstract}

\maketitle

\date{\today}

Is it possible to consider the relaxation of a closed 
quantum system to an apparently equilibrated state? 
In contrast to the deep understanding
we have of equilibrium quantum
statistical mechanics, non-equilibrium relaxation processes
are far from being fully understood. 
Specifically, the question of how Gibbs or relaxed states emerge 
dynamically is a question that is receiving a lot of attention recently
[1--9].
Part of the reason for the renaissance in the study of 
questions of non-equilibrium dynamics of quantum
many-body systems stems from the fact that 
systems have become available that 
promise to make such issues amenable to 
experiment \cite{Exp}: The states of cold atoms in optical lattices 
can be manipulated 
with a high degree of control, offering a testbed for
questions of non-equilibrium dynamics. 

The setting of interest in this work is the one of a {\it sudden quench}
[1--9, 11, 12].
Starting in the ground state of a local many-body Hamiltonian,
system parameters are suddenly changed such that
the old state is no longer an eigenstate of the new 
Hamiltonian, generating a non-equilibrium situation. The 
dynamics of the system is then monitored
in time. It has been conjectured that in such a non-equilibrium situation,
one may---in some sense---arrive at the 
 maximum entropy state  consistent with the expectation 
 values of the constants of motion fixed by the initial 
 state \cite{Rigol}, also referred to as a 
generalized Gibbs ensemble \cite{Jaynes,Page}. This is
appealing as it parallels
Jaynes' approach to equilibrium statistical mechanics.
Yet, of course, if the system can be meaningfully
treated as a {\em closed}
quantum system, one cannot expect the 
entire system to relax, as an initially pure state
will remain so in time [2--4]. The
entire information of the initial condition is still
present in the system, albeit in a very dilute fashion.

This is, however, by no means inconsistent with the 
expectation that the system may {\it locally} appear to 
be relaxed [2--5]. 
Locally, such a relaxation can well be true: 
Any subsystem may appear to be in a maximum entropy
state under the constraints dictated by the constants
of motion, and remain so for an arbitrary amount of 
time. Ref.\ \cite{Relax} introduces an instance in which 
this local relaxation of subsystems, referred to as the
{\it local relaxation conjecture}, can actually 
be proven rigorously to hold: When quenching
a state in a Mott phase in a Bose-Hubbard system
to the non-interacting deep superfluid phase, the 
reduced state of consecutive sites converges
(in trace-norm) to a maximum entropy state
consistent with the constants of 
motion \cite{Relax}---without having to invoke 
a time average. The intuition is that 
the quench creates local excitations that travel through the
lattice at a finite speed \cite{LR,Calabrese,Relax}. 
Their incommensurate influence then leads to a 
relaxation without environment. This intuition is corroborated 
in Ref.\ \cite{Barthel}, where local relaxation
for Gaussian initial states is shown.

\begin{figure}
\includegraphics[width=5.5cm]{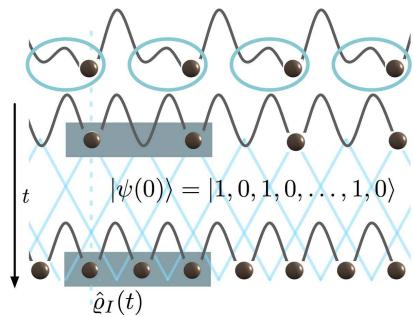}
\caption{\label{sketch}
Sketch of a local relaxation starting from
an initial condition of bosonic atoms being
present or absent, in even and odd
sites of a one-dimensional Bose-Hubbard system, 
achieved by imposing a superlattice to an optical lattice.}
\end{figure}

Unfortunately, while many situations that give rise to 
local relaxation may be generated, a challenge so far unresolved is 
to actually {\it probe signatures of local relaxation}: Demonstrating
local relaxation appears to necessitate local addressing,
a requirement that poses a great experimental
challenge in systems of atoms in optical lattices.

This dilemma will be resolved in this work, employing 
a simple yet promising idea: We will make use of a 
periodic setting; but we are by
no means obliged to stick to period $1$: The idea
is that we can well make use of a {\it period-$2$ setting}, 
exploiting optical superlattices. 
This approach will open up a way to quantitatively
explore local relaxation effects in experiment, without
the need of addressing single sites at any point. The
period-$2$ will allow for observing most of the relevant
signatures. To demonstrate the validity of this idea,
we will make use of analytical as well as numerical
methods, based on
a time-dependent density-matrix 
renormalization-group (t-DMRG) approach.
Our findings provide a simple 
guideline to what is to be expected in realistic 
experimental situations.

{\it Proposed experimental setup. --}
Ultracold atoms in optical lattices provide a great deal of
control over the system's parameters \cite{Exp}. In particular,
sudden quenches of parameters are accessible, and
on experimental timescales, systems can be treated as
essentially closed. To address the local detection problem, 
we propose to study local relaxation using {\em optical superlattices} 
\cite{Wells}. We follow the setup recently  realized 
by Bloch and coworkers  \cite{Wells},  considering
bosonic $^{87}$Rb atoms in a period-$2$ optical 
superlattice geometry. This experimental setting allows
for changing the relative intensity of the two optical lattices, shifting their relative position, 
and coupling and uncoupling double well 
potentials. In such double well potentials, one may 
introduce an alternating bias between the chemical potentials 
of neighboring sites.  This allows for the following three steps:

(I)  {\it Periodic patterns} of atoms can now be 
{\it prepared} by isolating double wells and introducing a 
bias between odd and even sites.
Further experimental techniques make sure that multiply
occupancies are highly suppressed, leaving, e.g., 
a sequence of empty and singly-occupied sites.

(II)  {\em Period-$2$ local} density
measurements can be performed 
by mapping odd and even sites to different Brillouin zones: 
Each part of a decoupled 
double well has multiple bands separated by well-defined 
energies. Biasing the odd sites relative to the even ones 
by an energy in excess of the separation energy of the band-separation energy, odd-site particles are reloaded into the higher bands of the even sites, whereas the even-site particles stay in the lowest band. A time-of-flight mapping then reveals 
the even-site particles in the first, the odd-site particles 
in the higher Brillouin zones.

(III)  {\em Correlations between even and odd sites} can
be measured using more sophisticated 
techniques \cite{Wells}.

{\it Setting and initial condition. --}
We propose to start from a two-periodic initial state 
prepared by the superlattice setup, where the odd sites
are occupied by exactly a single boson, all even
sites being empty, such that the initial state vector is
	$|\psi(0)\rangle=|1,0,1,0,\dots1,0\rangle$.
The $2a$-lattice is then suddenly
switched off, generating a quenched, non-equilibrium
situation, and the state vector  will evolve in 
time $|\psi(t)\rangle=\me^{-\mi t\hat H/\hbar} |\psi(0)\rangle$
according to the Bose-Hubbard Hamiltonian
\begin{equation*}
	\hat{H}=-J\sum_{i=1}^L
	(\hat{b}^\dagger_{i+1}\hat{b}_i+ 
	\hat{b}^\dagger_{i}\hat{b}_{i+1}) +
	\frac{U}{2}\sum_{i=1}^L\hat{n}_i\left(\hat{n}_i-1\right)-
	\mu\sum_{i=1}^L\hat{n}_i,
\end{equation*} 
$U$ and $J$ being the interaction 
and hopping parameters of the Bose-Hubbard model 
that can be calculated from the lattice 
parameters ($J=\hbar=1$). 
$L$ defines the system size, which will be taken to be even.  Occupation of higher order bands 
will not be considered, and
we stay within the limit of applicability of 
the Bose-Hubbard model. 
We will now see how local relaxation--- accompanied by
homogenization of densities and entanglement---manifests itself in 
such a periodic setting. 

\begin{figure}
\includegraphics[width=8.2cm]{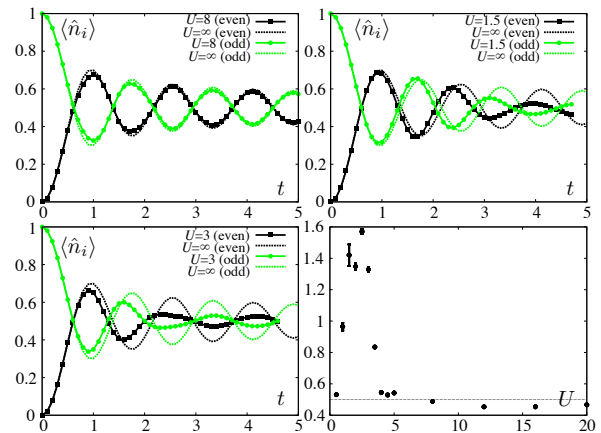}
\caption{\label{F1}Local density $\langle \hat{n}_i (t) \rangle$ vs.\ time, 
showing local relaxation. Shown is the time evolution of an even and 
an odd site for $U=0$ and $U=1.5$, $U=3$, and $U=8$, as well
as the estimated negative
exponents of asymptotic power-law decay (right lower figure).
Note the strong deviation from the non-interacting limit and the strong 
suppression of density oscillations for $U=3$, as well as the
similarity for $U=8$ with the limit $U=\infty$. }
\end{figure}

{\it Signatures of local relaxation for observable quantities. --}
We find that quantitative  
signatures of local relaxation can be obserbed 
by the global measurement  of the total occupation 
of even and odd sites (II), $\langle\hat{N}_{o,e}(t)\rangle$, 
and correlations between nearest neighbors 
(III), $\langle\hat{b}_i(t)\hat{b}_{i+1}(t)\rangle$. 
In a translationally invariant setting, such measurements
amount to accessing local observables as 
	$\langle\hat{N}_e(t) \rangle=
	\sum_{i=1}^{L/2}\langle\hat{n}_{2i}(t)\rangle
	=L\langle\hat{n}_{2i}(t)\rangle/2$.
The limiting cases $U=0$ and 
$U=\infty$ are or can be mapped to free models.
For $U=0$, one finds in the Heisenberg picture	
$\hat{b}_i (t) = \sum_{j=1}^L V_{i-j}(t)
	\hat{b}_j (0)$, where
	$V_{l}(t) = \sum_{k=1}^L\me^{-\mi t\lambda_k} 
	\me^{2\pi\mi k l /L }/L$ and 
	$\lambda_k=-\mu-2\cos(2\pi k/L)$.
In the limit $U\rightarrow\infty$, the interaction manifests 
itself in that bosons become hardcore, rendering
them solvable via the Jordan-Wigner transformation
to spinless fermions.
%$\hat{b}_n = \me^{-\mi \pi \sum_{m<n} \hat{c}^\dagger_m \hat{c}_m} %\hat{c}_n $.

To study the relaxation dynamics for finite $U$, we 
turn to the time-dependent variant 
\cite{tDMRG} of DMRG \cite{White}, allowing 
to follow the coherent time-evolution of strongly 
interacting quantum systems very precisely,
keeping up to $5000$ states in the
matrix-product state calculations.

In all of the considered cases, we do find local relaxation:
The intuition is that the incommensurate influences of 
travelling excitations \cite{Relax,Calabrese} 
lead to a mixing in time, and hence
the emergence of properties that locally appear like 
ones of maximum entropy states. These excitations proparate
at most with the Lieb-Robinson velocity, giving rise to an
approximate locality in the lattice \cite{LR}. The situation is 
especially clear for $U=0$, where relaxation is due to 
dephasing in the sense that freely propagating excitations 
lead to reduced state contributions of quickly oscillating phases 
that average out  \cite{Relax}. One can indeed 
prove that for any subblock $I$ of consecutive sites, the system
relaxes to 
the reduction of a  maximum entropy state given the 
constraints of motion.  This is true without time average,
arbitrarily exactly to any small error in trace-norm, 
and for an arbitrarily long time \cite{RelaxProof}. For 
interacting cases with $U>0$, we
would expect that for very short times, observables  
evolve as in the $U=0$ limit, with a crossover in 
behavior for longer time 
scales when they start interacting. We will explore to what
interaction strengths the remarkably simple
limiting pictures remain essentially 
valid and in what regimes one can identify genuinely different 
relaxation dynamics. 

{\it Time evolution of densities. --} Signatures of 
non-equilibrium relaxation dynamics are specifically
apparent in the observation of local densities.
Local densities evolve in both 
exactly solvable cases $U=0,\infty$ as
\begin{equation*}
	\left\langle \hat{n}_i (t)  \right \rangle 
	=\frac{1}{2}-\frac{(-1)^i}{2L}\sum_{k=1}^L
	\me^{4\mi t\cos(2\pi k/L)} 
	\rightarrow \frac{1}{2}-\frac{(-1)^i}{2}J_{0} (4t) 
\end{equation*}
for $L\rightarrow\infty$, where $J_n$ denote the
Bessel functions of first kind. 
Odd- and even-site densities relax symmetrically about 
the $n=1/2$ axis to $n=1/2$, with an asymptotic decay in time 
as $ t^{-1/2}+ o( t^{-1/2})$, see Fig.\ \ref{F1}.

For the interacting cases, all t-DMRG results are compatible with a 
relaxation of densities to $n=1/2$. As excitations are responsible for
local relaxation, on very short time scales ($t < 1$) 
particles have typically
not interacted yet and are not quite sensitive to different
values of $U$, giving rise to a similar behavior as for $U=0,\infty$.
The dynamics of relaxation deviates quite strongly for intermediate times, however.  
This is clearly exhibited in Fig.\ \ref{F1} (for $L=32$). 
The decay behavior as estimated from the data is also shown in 
by power-law exponents compatible with the data. 
We clearly see that for very small $U$ and $U>4$, the limiting 
non-interacting cases provide a very 
good approximation to the interacting dynamics
(one encounters a slope similar to $-1/2$, as for the limiting cases). 
For intermediate values of $U$, close to the critical point,
scattering appears to be most effective, leading to a strongly enhanced  damping and relaxation.
Such strong deviations from the limiting behavior are expected 
be visible experimentally, as a signature of interaction effects.

\begin{figure}
\includegraphics[width=8.5cm]{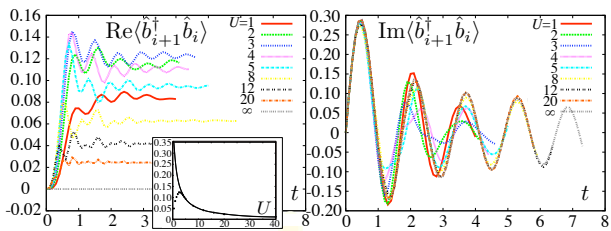}
\caption{\label{realpartnn} Real and imaginary 
parts of the correlations to neighbors $\langle \hat b_{i+1}^\dagger (t)
\hat b_i (t) \rangle$ as a function of time, for different values of $U$.
The inset shows the equilibrated  value of the real part of correlations to 
neighbors for large $t$ for different values of $U$ (with a solid line
proportional to $1/U$ as
a guide to the eye).}
\end{figure}

{\it Time evolution of correlators. -- } We now turn to 
nearest-neighbor correlator $\langle \hat{b}^\dagger_{i+1}(t) \hat{b}_{i} (t)  \rangle$. 
This quantity is specifically interesting, going beyond local densities: The build-up of correlations in time starting from the uncorrelated initial 
state becomes apparent. 
In the limiting free cases, we find
\begin{eqnarray*} 
	\langle \hat{b}^\dagger_{i+1}(t) \hat{b}_{i} (t) \rangle 
	=\frac{(-1)^{i}}{2L}\sum_{k=1}^L
	\frac{\me^{4t\mi\cos(2\pi k/L)}}{\me^{2\pi\mi k /L} } 
	\rightarrow
	- \frac{(-1)^{i}}{2\mi}J_{1} (4t),
\end{eqnarray*}
The real part of the correlator is zero for all times, whereas the 
imaginary part relaxes to $0$ with an asymptotics of 
$t^{-1/2}$, following a quick growth to a maximal value of 
about $0.28$ at time $t \sim 1/2$, reflecting
the buildup of correlations due to particle motion with speed approximately
linear in $J$. 
 
Fig.\  \ref{realpartnn} shows the real and imaginary parts of 
the correlators. The buildup of the imaginary 
part of correlations is largely independent of $U$, reflecting the 
fact that over the distance $1$ between particles at $t=0$,
few collisions have yet happened. 
When the interaction becoming visible, the relaxation dynamics
follows quite different paths, local relaxation again being fastest around $U \sim 3$.

For all finite $U$, the real part converges to a finite value. Indeed, 
for large $U$ the converged value is well-approximated by a $U^{-1}$ curve (for $U>4$).  This dependence is, in fact, exactly what one 
would expect in the thermal or Gibbs state of the Bose-Hubbard model. This can already be seen using
thermal perturbation theory, from which 
one obtains
\begin{eqnarray}
	\langle \hat b_{i+1}^\dagger \hat b_{i}\rangle = 
	\frac{J}{U}\sum_{n,m}\me^{-\beta(E_n+E_m )}
	n(m+1)\frac{\me^{\beta U(n-m-1)}-1}{z^2(n-m-1)},
	\nonumber
\end{eqnarray}
up to $o(J/U)$, 
where $z= \sum_{n} \me^{-\beta E_n}$ and $E_n=Un(n-1)/2-\mu n$ are the 
local energies of the unperturbed Hamiltonian.
So, indeed, within the validity of perturbation theory, we do find the anticipated 
 linear dependence on $1/U$, as seen also in DMRG simulations. 
This insight further corroborates the intuition that locally, 
the state is indistinguishable from the situation as if  the system was, globally, in a state maximizing the entropy,  respecting 
the constraints of motion \cite{Relax,Barthel}.

   {\it Entanglement dynamics. --} The travelling excitations
   give rise to entanglement between any subsystem and the 
   rest of the lattice  \cite{Entanglement,Dynamics}; 
   indeed to maximal entanglement
   given the constraints of motion if they are 
   local. The speed of information transfer also
   governs the dynamics at which bipartite and long-range entanglement
   is being built up. The buildup of entanglement
   is both a resource (being ultimately responsible for local relaxation) and
   a burden:  Linear entanglement growth as found here leads to an 
   exponential growth in the numerical resources, limiting 
   simulations to $t \sim 6 J$.
  
{\it Time scales of local relaxation. --} In all quantities studied so far,
one can identify three regimes in time:

(a) Initially, {\it correlations are being built up}. In this regime,
$Jt<1$,
the dynamics is largely independent of $U$, as 
collisions have not yet become important. 

(b) The second time regime is the one of actual {\it local relaxation}. The fast
oscillatory dynamics between neighboring sites is accompanied by
slow local relaxation, resulting from the incoming excitations from 
farther and farther sites, broadened by dispersion. This results in relaxation,
not due to decoherence but due to the dilution of information over the lattice. 
This information propagation is expected to
happen at a finite speed  related to $J$  \cite{LR},
in a ``ballistic transport''. In the free models, one finds a polynomial decay: The Bessel function fulfills 
$J_0(x) = x^{-1/2}+ o(x^{-1/2})$. For finite $U$, 
numerics is consistent with 
polynomial decay, which for intermediate $U$ 
seems to be much {\it faster}.
 
(c) The third and very large time, not yet visible on the simulation time scales, 
is the  {\it recurrence time} where the finite size of the quantum system becomes visible.

The interaction strength also marks three regimes:

(A) For small interactions $U<1$
the dynamics strongly resembles the
non-interacting bosonic limit.

(B) For large values of $U$ the dynamics is very similar to the 
hardcore bosonic or free fermionic limit, with relaxation exponents
being similar to $-1/2$. The observed local 
correlations proportional to $1/U$
are consistent with assuming that we locally observe a global
state having maximum entropy.  

(C) 
The case of intermediate $U \sim 3$ appears to mark the crossover between the two free cases of $U=0$ and $U=\infty$, 
characterized by the most efficient relaxation.

{\it Quasi-momentum distribution. --} Remarkably, the
quasi-momentum distribution (QMD), measurable  via time-of-flight, 
defined as $S(q,t) =  \sum_{i,j=1}^L 
	\me^{\mi q(i-j)} \langle \hat{b}^\dagger_i(t) \hat
	b_j (t) \rangle/L$ can be shown not to 
relax for $U=0$. This creates,
for small $U$, the interesting situation that local
relaxation can be probed, while at the same time signatures
of the memory of the initial condition can be certified by 
the absence of relaxation for the QMD \cite{LongTimes}. 
      
{\it Effect of a harmonic confinement potential. --} 
We find little influence on the local quantities, up to the finite-size recurrence time, 
which can be shortened in the presence of a realistic trap: the generated excitations no longer travel with a constant 
speed, but are slowly reflected.
      
{\it Summary. --} In this work, we have introduced a setting
allowing for the observation of apparent thermalization in closed
quantum many-body systems, without the need of addressing 
single sites of a lattice system.
We show how, making use of optical superlattices, 
signatures of
local relaxation, including relevant time scales and dependencies on
interactions, can be probed. In such dynamics, 
local maximum entropy states---Gibbs states in the 
absence of further constraints---emerge without 
having any external heat bath: Instead, in a symmetric 
fashion, each system forms the effective environment of the other
in complex many-body dynamics. 
It is the hope that ideas along the line of the present work 
open up a way of experimentally quantitatively
simulating an instance of the fundamental physical process of 
thermalization. 
     
{\it Acknowledgements. --} We thank
the EU (QAP, COMPAS), the EPSRC, the EURYI, 
and Microsoft Research for support, and 
T.\ Barthel,  I.\ Bloch, 
C.\ Kollath, M.B.\ Hastings, and
T.J.\ Osborne for discussions.


\begin{thebibliography}{99}
\bibitem{RelaxProof}
	For any $\varepsilon>0$, $t_{\rm rec}>0$ there exists a 
	system size $L$
	and a relaxation time $t_{\rm rel}>0$ such that the 
	reduced state satisfies 
	$\|\hat \varrho_I(t)- \hat \varrho_{\rm max}\|_1<\varepsilon$
	for all $t\in [t_{\rm rel},t_{\rm rec}]$  
	\cite{Relax}.
\bibitem{LongTimes}
	For very long times and considerable
	$U$, one may find a relaxation in momentum
	space due to coupling of eigenmodes, as being 
	corroborated by flow approaches \cite{Kehrein}.
\end{thebibliography}
\end{document}